\documentclass[10pt]{article} 
\pdfoutput=1
\usepackage[utf8]{inputenc}
\usepackage[T1]{fontenc}
\usepackage[english]{babel}
\usepackage{geometry} 
\usepackage{authblk}
\usepackage{amsthm}
\usepackage{csquotes}
\usepackage{amsmath}
\usepackage{amsfonts}
\usepackage{amssymb}
\usepackage{bbold}
\usepackage[backend=biber,
                style=nature,
                sorting=none]{biblatex}
\usepackage[labelfont=bf]{caption}
\usepackage{changepage}
\usepackage{float}
\usepackage{graphicx}
\usepackage{mathrsfs}
\usepackage{mathtools}
\usepackage[arrowdel]{physics}
\usepackage{soul}
\usepackage{xcolor}
\usepackage{hyperref}
\hypersetup{colorlinks,allcolors=black}

\DeclareUnicodeCharacter{03BC}{\ensuremath{\mu}}

\title{\textbf{Time-resolved spectroscopy of \\ noise-driven collective states of light}} 
\author{Diego Piciocchi$^{1,\dagger,*}$, Ina Heckelmann$^{1,\dagger,*}$, Alexander Dikopoltsev$^{1}$, \\ Michael Schreiber$^{2}$, Mathieu Bertrand$^{1}$, Mattias Beck$^{1}$, \\ Christian Jirauschek$^{2}$, Oded Zilberberg$^{3}$, Jérôme Faist$^{*,}$}

\affil[1]{Institute for Quantum Electronics and Quantum Center, ETH Zürich, 8093 Zürich, Switzerland.}
\affil[2]{TUM School of Computation, Information and Technology, Technical University of Munich (TUM), Garching, D-85748, Germany.}
\affil[3]{Department of Physics, University of Konstanz, Konstanz, 78464, Germany.}
\affil[*]{\rm{Corresponding authors. Email: dpiciocchi@phys.ethz.ch, iheckelmann@phys.ethz.ch, jfaist@phys.ethz.ch}}
\affil[$\dagger$]{\rm{These authors contributed equally to this work.}}

\date{}

\begin{document}
  \maketitle

\begin{abstract}
We study a collective liquid state of light in a fast-gain laser. Controlled temporal noise on the cavity modulation creates a fluctuating linear potential along the synthetic frequency lattice of the cavity modes. We identify three regimes of lattice occupation as noise increases: an extended distribution, a Gaussian envelope, and exponential localization. Time-resolved spectroscopy on single realizations of noise reveals distinct dynamics in the latter two: transport persists in the Gaussian regime, modulated by the fluctuating potential, but is fully suppressed at all times in the localized regime. Averaging over many noise realizations shows that noise reduces the transport speed and confirms ergodicity of the system.
\end{abstract}

\section{Introduction}

The study of random processes ubiquitously shapes our understanding of dynamics in physical systems~\cite{parisi_nobel_2023}. From the role of random forces in Brownian motion~\cite{einstein_uber_1905}, stochastic descriptions have permeated diverse disciplines, explaining fluctuations in electronic circuits~\cite{gardiner_stochastic_2009}, chemical reaction rates~\cite{kramers_brownian_1940}, option pricing~\cite{paul_stochastic_2013}, population diffusion~\cite{okubo_diffusion_2001}, and even collective decision making~\cite{frank_deterministic_2009, bogacz_physics_2006}.

The interplay between noise and nonlinearity, inherent to a wide range of systems, is surprisingly complex~\cite{moss_noise_1989}, since noise can alter the response of a nonlinear system. Sufficiently strong noise can introduce underlying disorder, which is known to impact transport phenomena~\cite{anderson_absence_1958, sacha_anderson_2016, apffel_experimental_2022}. This is particularly evident in non-equilibrium systems, like those driven to counteract dissipation~\cite{dykman_fluctuational_1998, eichler_classical_2023}. In such systems, noise can counterintuitively induce order, triggering phase transitions~\cite{horsthemke_noise-induced_2006} and switching between stationary states~\cite{heugel_proliferation_2024}. Striking examples include stochastic and coherence resonances where noise, in combination with a periodic drive, enables the detection of weak signals in systems ranging from nanomechanical oscillators~\cite{badzey_coherent_2005} and SQUIDs~\cite{wiesenfeld_stochastic_1995} to neuronal firing~\cite{mori_noise-induced_2002} and climate variability~\cite{benzi_stochastic_1982}.

Optical systems provide an ideal platform to investigate the interplay of noise and nonlinearity, due to their inherent controllability and the long coherence time of light. These features have enabled the realization of diverse systems to study light dynamics in ordered and disordered environments, demonstrating lattice transport~\cite{morandotti_experimental_1999, christodoulides_discretizing_2003} and its halt in disordered structures~\cite{schwartz_transport_2007, lahini_anderson_2008, yu_engineered_2021} that can even be induced by virtual scattering processes~\cite{dikopoltsev_observation_2022}. At the same time, topological photonics~\cite{hafezi_robust_2011, rechtsman_photonic_2013, ozawa_topological_2019} in real and synthetic dimensions~\cite{ozawa_synthetic_2016} has been used to mitigate the effects of disorder. Introducing nonlinearities to the propagation media~\cite{bloembergen_nonlinear_1965} leads to collective behaviors~\cite{carusotto_quantum_2013} such as condensation~\cite{kasprzak_boseeinstein_2006, klaers_boseeinstein_2010}, superfluidity~\cite{amo_superfluidity_2009} and supersolidity~\cite{trypogeorgos_emerging_2025}. In this context, the role of the drive—necessary to compensate for the intrinsic dissipation in optical systems—in stabilizing and controlling many-body states has attracted significant interest~\cite{ai_optically_2025, carusotto_how_2025}. Recent advancements have explored the use of nonlinear time-dependent optical materials to implement controlled temporal modulations, including random fluctuations~\cite{sharabi_disordered_2021, galiffi_photonics_2022, engheta_four-dimensional_2023}. 

Early investigations into the interplay of noise and nonlinearity in optical systems focused on instabilities and stochastic resonances in lasers~\cite{mcnamara_observation_1988, barbay_stochastic_2000}, revealing noise-induced transitions between lasing states. More recently, some works have highlighted the high degree of control achievable over collective multimode states of light in nonlinear systems driven out of equilibrium~\cite{piccardo_frequency_2020, kazakov_driven_2025}. This fueled the study of rich physics such as nonlinear quantum walks~\cite{heckelmann_quantum_2023} and Bloch oscillations~\cite{dikopoltsev_collective_2025, englebert_bloch_2023} dynamics in the frequency domain, light manipulation through synthetic magnetic fluxes~\cite{piciocchi_frequency_2025}, hybridized solitons~\cite{letsou_hybridized_2025}, and non-Hermitian topology~\cite{schneider_ultrafast_2025}. In these systems, inherent suppression of temporal fluctuations hold significant potential for structuring light states~\cite{lukashchuk_chaotic_2023, roy_liquid_2025}. Despite this extensive research on collective light dynamics in driven nonlinear optical systems, the effect of controlled noise injection on multimode states remains underexplored.

\begin{figure}[t]
    \centering
    \includegraphics[width=0.85\textwidth]{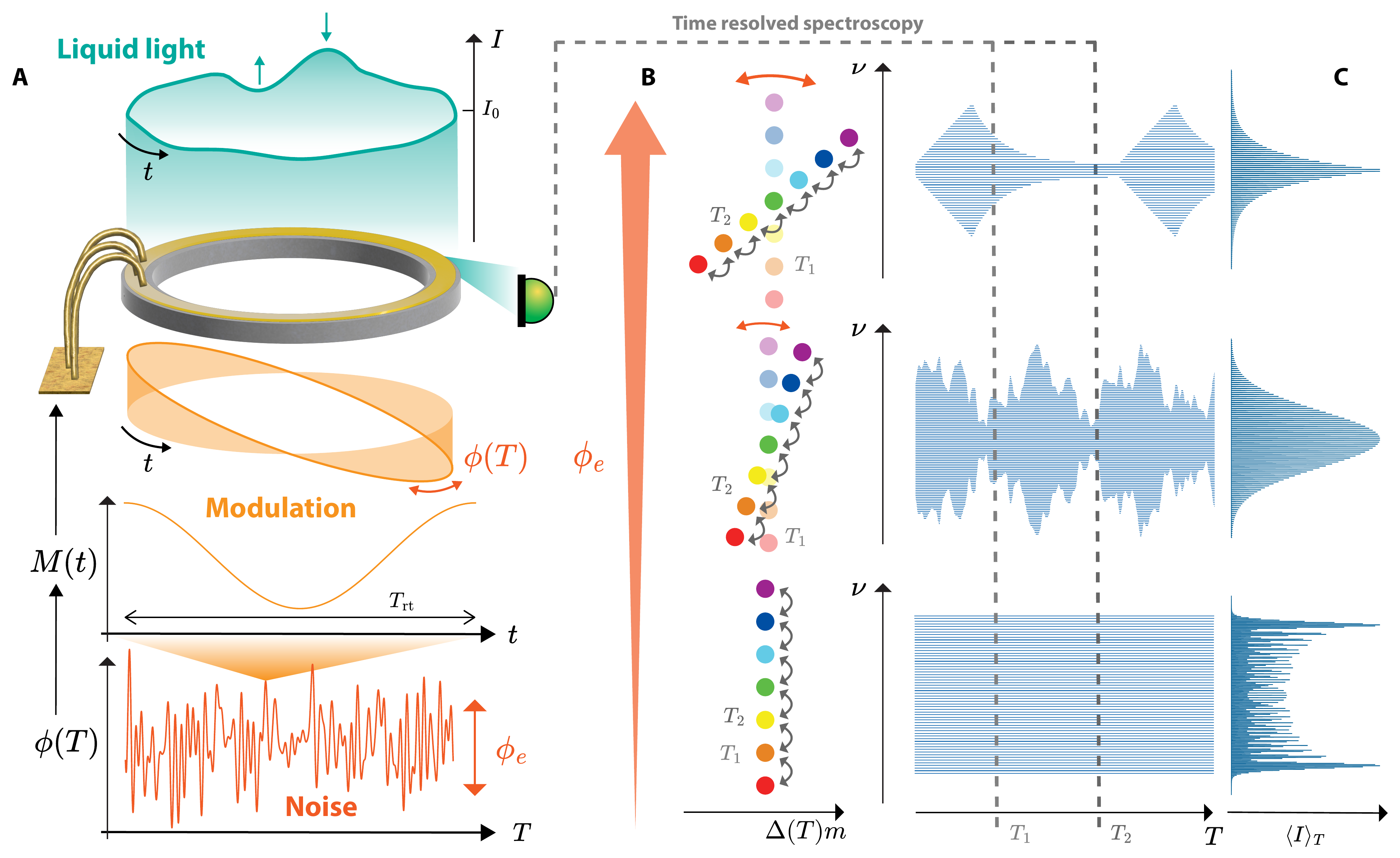}
    \caption{\textbf{Time-resolved spectroscopy of noise-driven liquid light.} (A) Ring-shaped Quantum Cascade laser, operating at a quasi-constant intensity $I_0$ (blue) due to the fast gain, which suppresses intensity fluctuations. The laser is subject to resonant modulation $M(t)$ (orange). Noise is injected in the system by phase modulation of the resonant drive $M(t)$ with a noisy signal $\phi(T)$ of controlled amplitude $\phi_e$ (red), fluctuating on a timescale much slower than the cavity round-trip $T_{\rm rt}$. Here, $t$ indicates real laboratory time, while $T$ is a slow time measured in units of $T_{\rm rt}$. (B) In the frequency ($\nu$) domain, each cavity mode maps to the $m$-th site of a one-dimensional lattice. The resonant modulation induces nearest-neighbor coupling (grey arrows) and the injection of noise maps to a fluctuating linear potential $\Delta(T)m$ in frequency space. (C) We measure the lattice occupation by performing time-resolved spectroscopy, measuring the spectrum at different times ($T_{1,2}$ in the figure being two examples), corresponding to different values of $\Delta$. Three different regimes are identified in the time evolution of the spectrum, shown in the figure with the blue time-resolved maps, each with a different envelope for the average spectrum (rightmost column).}
    \label{fig:concept}
\end{figure}

Here, we investigate the collective liquid state of light in a fast-gain laser under noise injection, and observe abrupt noise-induced localization transitions in both experiment and numerical simulations. To this end, we apply a resonant drive to the laser cavity, phase-modulated with controlled noise (Fig.~\ref{fig:concept}A). The laser's ultrafast gain suppresses intensity fluctuations on the fastest timescale, resulting in a quasi-constant intensity, analogous to an incompressible liquid surface~\cite{dikopoltsev_collective_2025} (Fig.~\ref{fig:concept}A). This constant intensity implies that the dynamical features arising from the modulation are imprinted in reciprocal space and become accessible through spectral measurements. In this space, each laser cavity mode corresponds to a site on a synthetic frequency lattice, and the injected noise maps to a linear potential fluctuating in time, whose strength is controlled by the noisy modulation of the drive (Fig.~\ref{fig:concept}B). Our time-resolved spectral measurements reveal a transition through three distinct regimes: a broad synthetic lattice occupation at low noise, a Gaussian envelope upon temporal averaging at intermediate noise, and exponential localization at high noise levels (Fig.~\ref{fig:concept}C). This observation of localization in synthetic space induced by temporal disorder demonstrates how controllable drive noise can shape the dynamical behavior of a collective light state, enhancing the understanding -and, potentially, the control- of the multimode emission dynamics of fast-gain lasers. 

\section{Injecting noise and mapping to the frequency domain}

Our experimental platform is a Quantum Cascade Laser (QCL~\cite{meng_dissipative_2021, heckelmann_quantum_2023}) with a circular cavity (Fig.~\ref{fig:concept}A), operating at a wavelength around 8$\mu$m. The smooth sidewalls minimize backscattering, allowing single-mode operation in either the clockwise (CW) or counterclockwise (CCW) direction~\cite{seitner_backscattering-induced_2024}. We modulate the laser's driving current $J(t)$ at the resonance frequency of the cavity, $f_{\rm res}\approx15.74$GHz, which induces a phase modulation to the intracavity field via the linewidth enhancement factor~\cite{opacak_spectrally_2021}. 

To inject noise, we apply phase modulation to the periodic drive using a noise signal (for details concerning the implementation, see~\cite{supmat}), resulting in a current $J(t)=J_0 + J_{\rm mod}\cos(2\pi f_{\rm rep} + \phi(t))$. Here, $J_0$ is the DC bias, $J_{\rm mod}$ is the AC modulation depth, and $\phi(t)=\phi_e\cdot N(t)$ is the random modulating phase. This phase is determined by a Gaussian noise signal $N(t)$ with an adjustable excursion $\phi_e$, which serves as our control parameter for the injected noise strength (Fig.~\ref{fig:concept}A). The electric field $E(t)$ inside the laser cavity, in the copropagating frame of the free-running lasing mode, is governed by the complex Ginzburg-Landau equation: 
\begin{equation}\label{eq:cGLE}
\begin{split}
     i\dv{E(t)}{t} = i\left[g_0\left(1-\frac{I(t)}{I_{\rm sat}}\right) -\alpha\right] E(t)  +\left(\frac{i}{2}g_c - \frac{\beta}{2}\right)\nabla^2 E(t) + 2C\cos\left[Kz - \phi_{\rm e}N(t)\right]E(t).
\end{split}
\end{equation}
where $g_0$ is the unsaturated gain, $I$ and $I_{\rm sat}$ are the intensity and saturation intensity, $\alpha$ is the total loss, $g_c$ is the gain curvature, $\beta$ is the dispersion, $2C$ is the phase modulation coefficient, $z$ is a spatial cavity coordinate and $K=\frac{2\pi}{L}$ with the cavity length $L\approx5.77$mm. 

As previously discussed and reflected in the governing equation, the ultrafast gain saturation ensures that the gain responds on timescales negligible compared to the intracavity dynamics. This effect suppresses intensity fluctuations, leading to the quasi-constant intensity characteristic of a collective liquid state~\cite{dikopoltsev_collective_2025}. Consequently, the system's dynamics must be observed in reciprocal space. We can describe our system in this space by expanding the field in terms of spatial modes of amplitude $A_n$: $E=\sum_n A_ne^{-inKz}$. Neglecting the nonlinear non-Hermitian term related to gain saturation, the system's Hamiltonian reads (see~\cite{supmat} for derivation):
\begin{equation}\label{eq:random}
    H = \sum_m \left[(D-i G)m^2 +\Delta(T) m\right] a_m a^\dagger_m + C \left[a^\dagger_{m-1}a_m + a^\dagger_{m+1}a_m\right],
\end{equation}
where $D=\beta K^2/2$, $G=g_c K^2 / 2$ and $\Delta(T)=\frac{\phi_{\rm e}}{2\pi f_{\rm res}} \dv{N(T)}{T}$. In the frequency domain, each cavity mode maps to a site on a synthetic lattice. Therefore, measuring the spectrum directly reveals the occupation of this lattice. In this framework, dispersion and gain curvature introduce a complex quadratic potential, while the periodic drive induces nearest-neighbor coupling with amplitude $C$. Notably, the injected noise maps to a randomly fluctuating linear component added to the potential for this synthetic lattice (Fig.~\ref{fig:concept}B).

\section{Impact of noise: distinct regimes}

To directly probe the occupation of the $m$-th lattice site, $\abs{A_m}^2$, we measure the laser's optical spectrum $S(\nu)$ using a Fourier Transform InfraRed (FTIR) spectrometer as a function of the phase excursion $\phi_{\rm e}$. The intensity of the $m$-th mode, with frequency $\nu_m$, is given by the time average $S(\nu_m)= \frac{1}{T_{int}} \int_0^{T_{int}} \abs{A_m(t)}^2 \text{d}t$, where $T_{int}$ is the measurement time (top panels, Fig.~\ref{fig:concept}C). Given that the measurement time for these spectra is $T_{int}\approx 1$s, much longer than the oscillation timescale of $\Delta(T)$ (on the $\approx\mu$s scale), these spectral measurements reflect the long-time average of the lattice occupation.
\begin{figure}[t]
    \centering
    \includegraphics[width=0.65\textwidth]{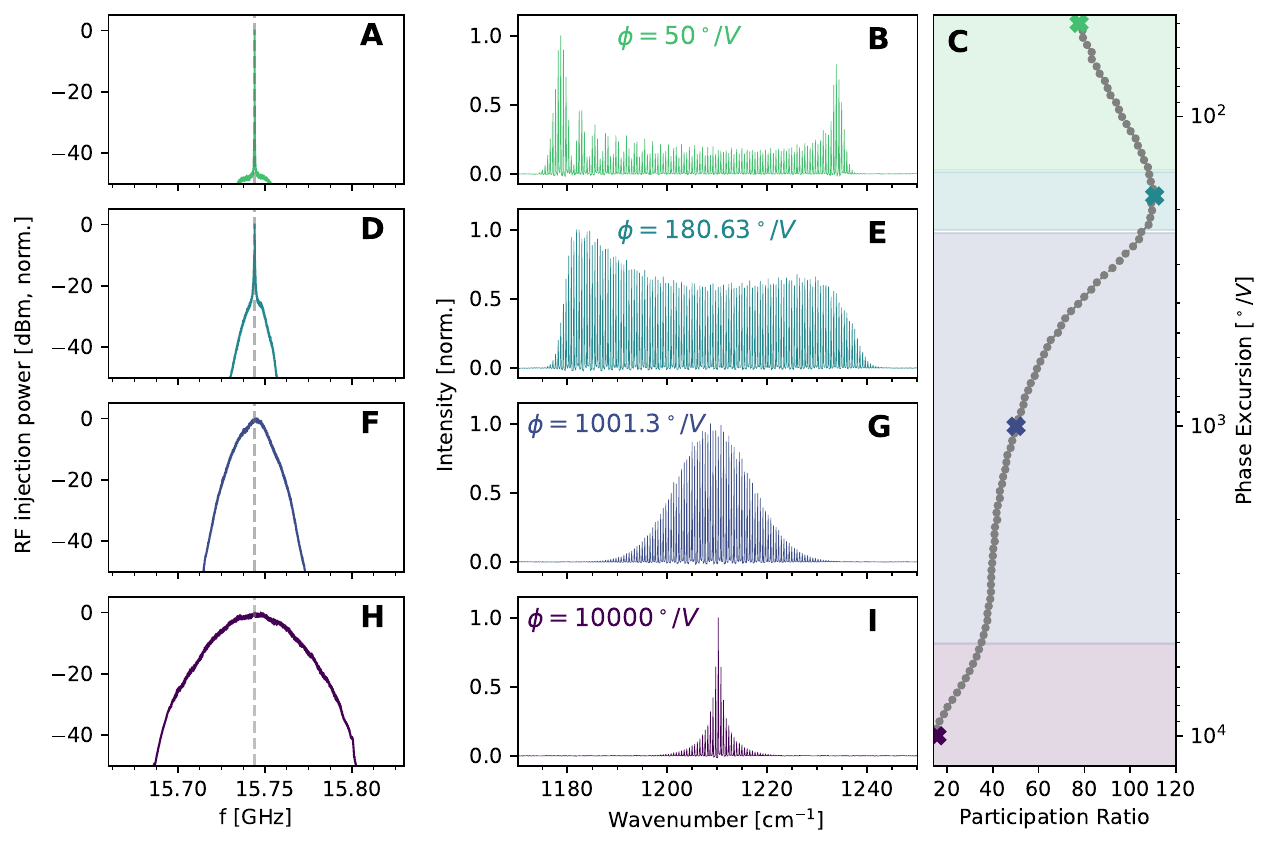}
    \caption{\textbf{Evidence of transitions induced by temporal noise.} (A,B) Electrical injection spectrum for low values of injected noise (A) and corresponding laser spectrum (B). (C) Participation ratio as a function of injected noise ($\phi_e$). Different regions are shaded with colors matching the corresponding electrical injection spectrum and laser spectrum. (D,E) Electrical injection spectrum for values of injected noise (D), corresponding to a flattened laser spectrum (E). (F,G) Electrical injection spectrum for intermediate values of injected noise (F), corresponding to a spectrum with Gaussian envelope. (H,I) Electrical injection spectrum for high values of injected noise (H), at which exponential localization is observed in the laser spectrum (I).}
    \label{fig:transition}
\end{figure}

When no noise is injected ($\phi_e = 0$), resulting in a narrow electrical injection spectrum (Fig.~\ref{fig:transition}A), the nonlinearity induces quantum random walk and locking dynamics in synthetic space, leading to a broad steady-state spectrum (Fig.~\ref{fig:transition}B), as demonstrated in a recent study~\cite{heckelmann_quantum_2023}. Fig.~\ref{fig:transition}C shows the participation ratio $\rm{PR} = \left(\sum_m I_m \right)^2 / \sum_m I_m^2$, where $I_m = S(\nu_m)$. As the injected noise increases (Fig.~\ref{fig:transition}D), the average spectrum progressively flattens (Fig.~\ref{fig:transition}E) and the participation ratio increases, reaching a maximum (Fig.~\ref{fig:transition}C). Upon further increasing of the injected noise (Fig.~\ref{fig:transition}F), the participation ratio trend reverses, and the spectral envelope becomes Gaussian (Fig.~\ref{fig:transition}G) with a width that decreases with increasing $\phi_{\rm e}$. Finally, at high $\phi_{\rm e}$ values (Fig.~\ref{fig:transition}H), a further abrupt change in the participation ratio curve occurs, and the spectrum becomes exponentially localized (Fig.~\ref{fig:transition}I). As detailed in~\cite{supmat}, we confirm the correspondence between the transitions in envelope shape and the changes in slope of the participation ratio (Fig.~\ref{fig:transition}C), by fitting the spectra using different envelope functions. Each shaded region in Fig.~\ref{fig:transition} indicates the region where the best fit corresponds to the shape of the example spectrum with the same color in panels B,E,G,I. Our observations provide evidence for a localization transition in a synthetic frequency space, driven by noise in a collective liquid-light state.   

\section{Time-resolved dynamics}

The results in Fig.~\ref{fig:transition} demonstrate that injected noise, resulting in temporal disorder for the synthetic lattice, induces localization. However, they leave the dynamical response of the collective liquid-light state unexplored. To address this, we developed a time-resolved spectroscopy technique that follows the system’s reaction to a single realization of the noise time trace. This enables real-time imaging of the evolving lattice occupation in synthetic space, avoiding averaging over multiple noise realizations, granting direct access to dynamical information otherwise inaccessible.

Starting from the single mode operation of the free-running QCL, we apply a sudden quench by initiating both bias current modulation and noise injection at $t=0$. Then, time-resolved spectroscopy (Fig.~\ref{fig:concept}C) directly reveals the time evolution of the synthetic lattice sites' occupation, $\abs{A_m(T)}^2$.
\begin{figure}[t]
  \begin{minipage}{0.49\textwidth}
    \centering
    \includegraphics[width=\linewidth]{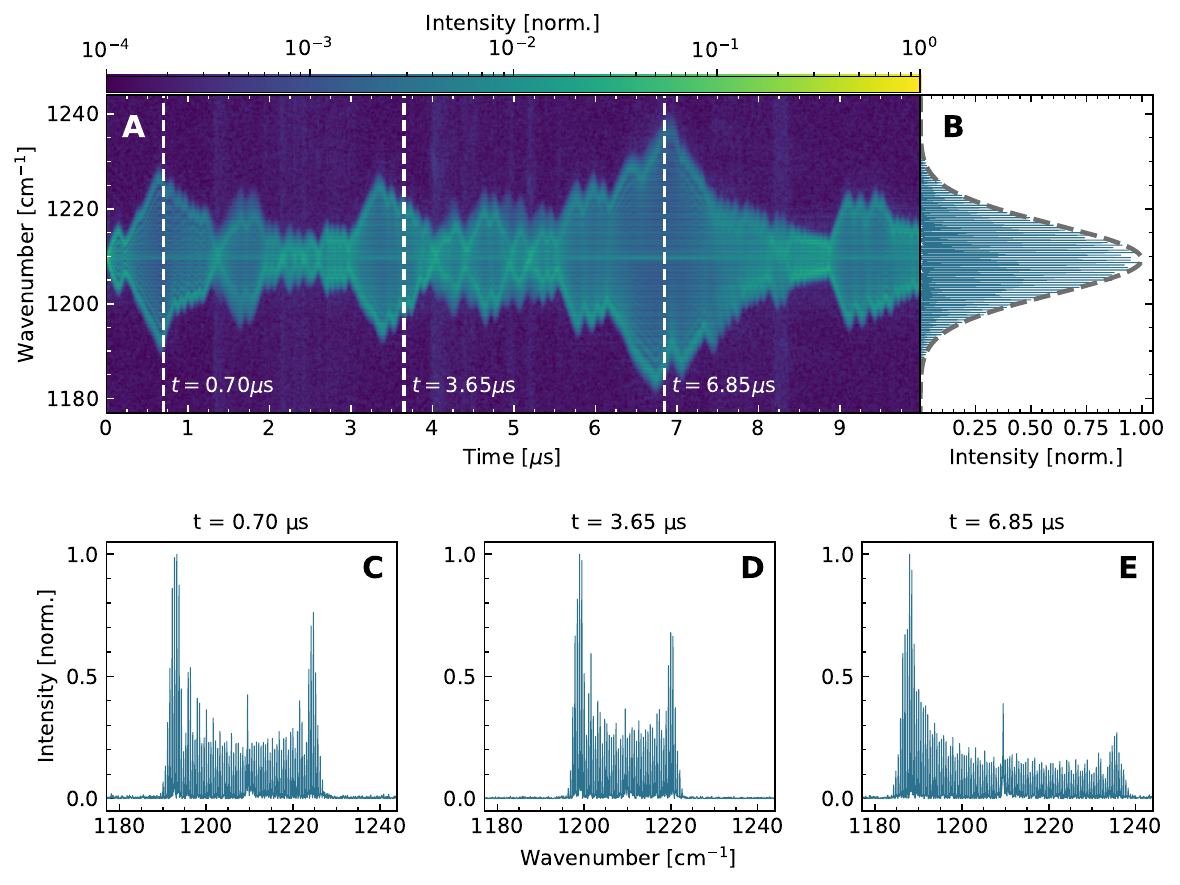}
    \caption{\textbf{Time-resolved spectroscopy in the Gaussian regime.} (A) Experimental time-evolution of the laser spectrum under noise injection with $\phi_e=1000^\circ/V$ (B) Average spectrum for the same value of the phase and Gaussian fit (dashed gray curve). (C-E) Individual spectra at times $T=0.7,3.65,6.85\mu$s.}
    \label{fig:gauss}
  \end{minipage}\hfill
  \begin{minipage}{0.49\textwidth}
    \centering
    \includegraphics[width=\linewidth]{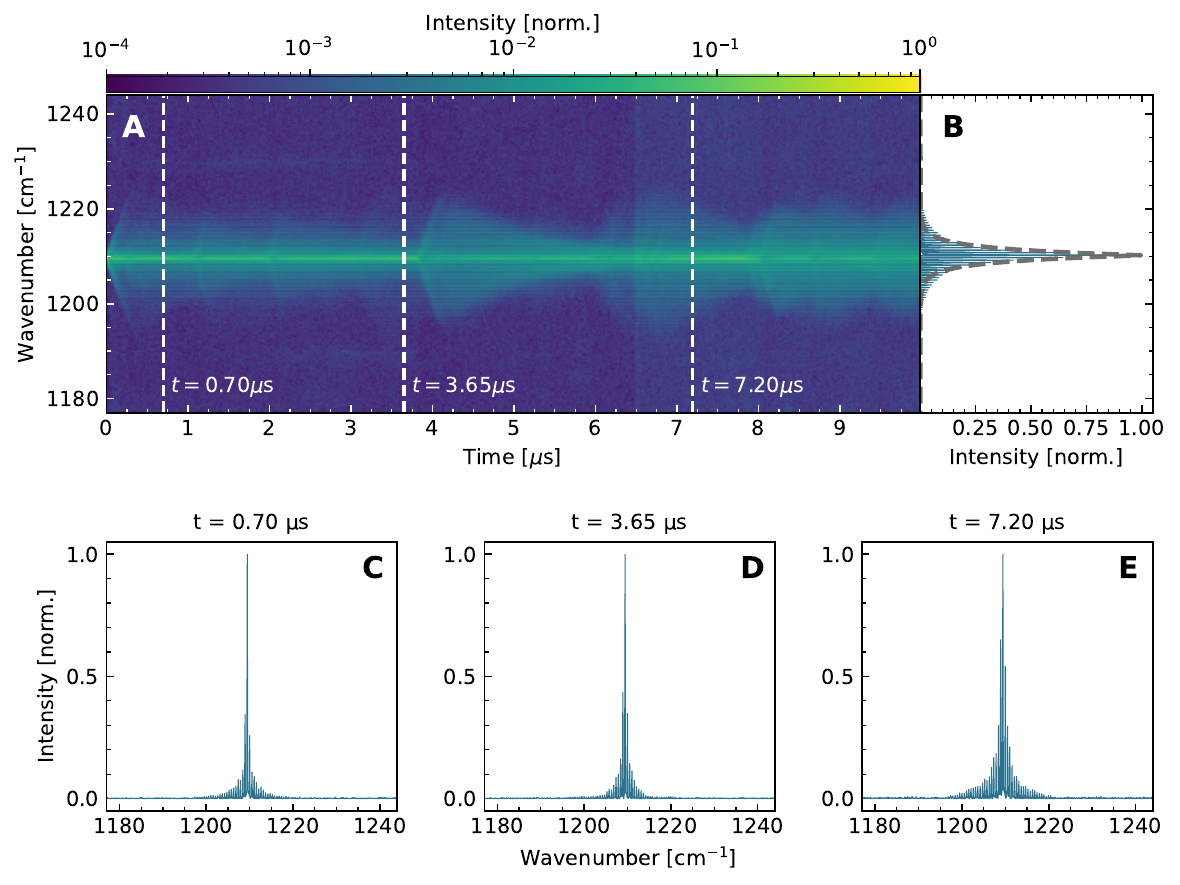}
    \caption{\textbf{Time-resolved spectroscopy in the localized regime.} (A) Experimental time-evolution of the laser spectrum under noise injection with $\phi_e=10000^\circ/V$ (B) Average spectrum for the same value of the phase and fit to an exponentially localized spectrum (dashed gray curve). (C-E) Individual spectra at times $T=0.7,3.65,7.20\mu$s.}
    \label{fig:exploc}
  \end{minipage}
\end{figure}
We measured this temporal evolution for two noise levels ($\phi_{\rm e}$), corresponding to regimes with Gaussian (Fig.~\ref{fig:gauss}) and exponentially localized (Fig.~\ref{fig:exploc}, see ~\cite{supmat} for experimental details) average occupations. Both Fig.~\ref{fig:gauss} and Fig.~\ref{fig:exploc} illustrate that the average spectra (Fig.~\ref{fig:gauss} B and Fig.~\ref{fig:exploc}B) arise from integrating the dynamic evolution over time, as the system does not reach a steady state within our observation window (see ~\cite{supmat} for details on the integration time). A key distinction is that while no single instantaneous spectrum matches the average Gaussian shape (Fig.~\ref{fig:gauss}C-E), all instantaneous spectra in the exponentially localized regime match the average exponentially localized form (Fig.~\ref{fig:exploc}C-E). In that case, all the spectra along the time trace match the functional form of the average spectrum. This suggests that in the Gaussian regime, the occupation across the synthetic lattice continues to evolve and can still potentially broaden (as in the absence of noise), whereas in the localized case, temporal disorder effectively halts transport along the synthetic lattice. This behavior -accessible only through time-resolved spectroscopy- reveals a fundamental distinction between the two regimes. In states with a Gaussian envelope, transport along the synthetic dimension remains possible and is controlled by the evolution of the linear component of the lattice potential. In contrast, in the localized regime, transport is strongly suppressed, resulting in localized states at all times.

Numerical simulations based on the cGLE model (Eq.~\ref{eq:cGLE} and ~\cite{supmat}) show good agreement with the previously discussed experimental results. They highlight that the observed distinct regimes can be accessed by solely and continuously tuning the phase excursion $\phi_e$. In addition, they underpin the importance of the fast gain recovery for the emergence of the discussed transitions (the effect of noise injection on a slow gain system is discussed in~\cite{supmat}). 

\begin{figure}[t]
    \centering
    \includegraphics[width=0.67\textwidth]{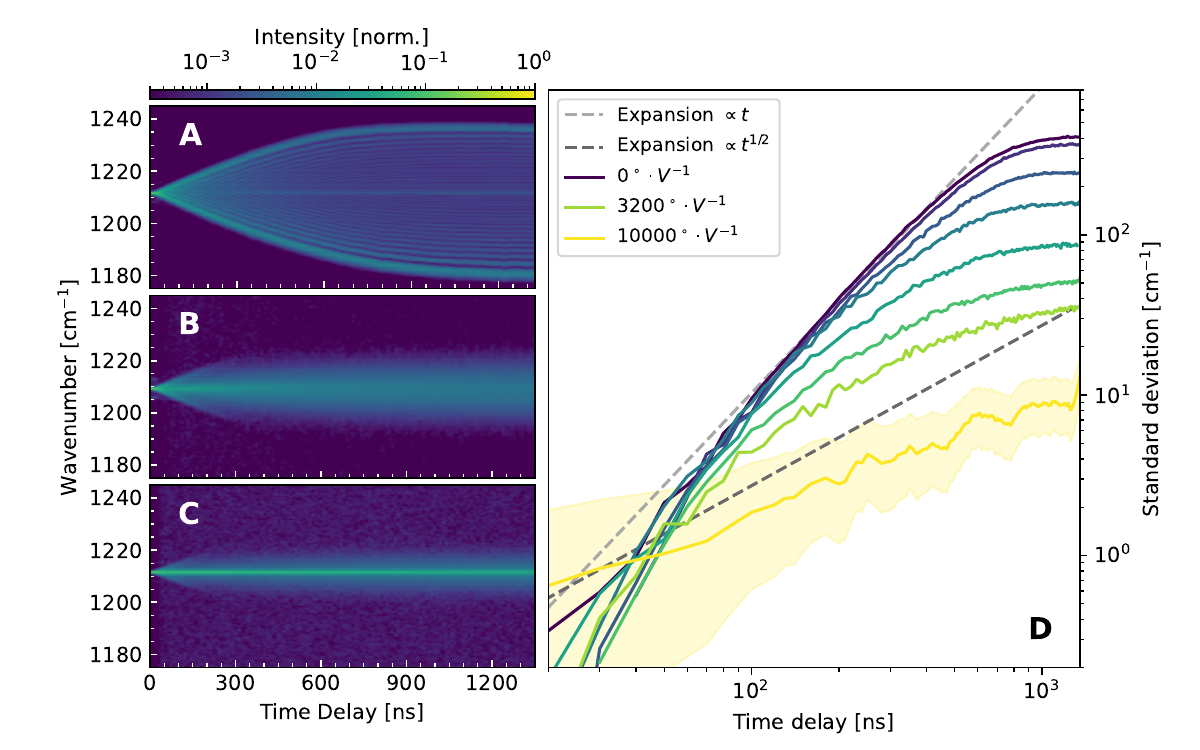}
    \caption{\textbf{Decrease in the rate of expansion.} (A-C) Time-resolved lattice occupation, averaged over many different traces of the applied force with the same standard deviation, for $\phi_e=0^\circ/V$ (A), $\phi_e=1000^\circ/V$ (B) and $\phi_e=10000^\circ/V$ (C). (D) Expansion rate under the effect of a given noise level extracted from the spectral expansion. }
    \label{fig:rates}
\end{figure}

The measurements in Fig.~\ref{fig:gauss} and Fig.~\ref{fig:exploc} depict the system's dynamics under noise injection, observed in reciprocal space as the synthetic lattice occupation. The noise strength is determined by the phase excursion $\phi_{\rm e}$, which is the relevant parameter, independent of a specific noise realization's temporal details. Thus, examining the impact of $\phi_{\rm e}$ alone on the system's dynamics is crucial. To isolate the effect of $\phi_{\rm e}$ we averaged the time evolution obtained from multiple different noise traces of the same duration and scaling (see ~\cite{supmat} for experimental details). The resulting averaged spectral dynamics for several $\phi_{\rm e}$ values are shown in Fig.~\ref{fig:rates}A-C.

We determined the spectral expansion rate from the increasing spectral bandwidth (Fig.~\ref{fig:rates}D). Starting from the single mode at $t=0$, the modulation initially leads to broadening of the spectrum. Without noise, this broadening is linear in time (~\cite{heckelmann_quantum_2023}, Fig.~\ref{fig:rates}A), indicating ballistic transport ($\sigma\propto t$, Fig.~\ref{fig:rates}D) along the synthetic lattice. This broadening eventually stabilizes at a certain bandwidth~\cite{heckelmann_quantum_2023}. Therefore, we only focus on the early-time expansion to distinguish the regimes (see ~\cite{supmat} for details). 

As the injected noise increases, the spectral expansion rate decreases, transitioning through the Gaussian regime towards a rate of $\sigma\propto t^{1/2}$, a signature of classical Random Walks. At very high noise levels, the system becomes confined to a few modes with exponentially decaying amplitudes, signifying inhibited transport, also reflected in a strongly reduced sub-diffusive expansion rate ($\sigma\approx \text{const.}$, Fig.~\ref{fig:rates}D).

Furthermore, regardless of the noise strength, all instantaneous spectra along the time trace exhibit the functional form of the average spectrum for that regime (quantum walk-like, Gaussian, or exponentially localized). This confirms that time-averaging within a single noise realization (as in Fig.~\ref{fig:gauss} and Fig.~\ref{fig:exploc}) is equivalent to averaging the spectrum at a given time $t^\star$ over many noise realizations between $t=0$ and $t=t^\star$, confirming the ergodicity of the system.

\section{Concluding remarks}

In summary, we have experimentally and numerically demonstrated localization transitions along a synthetic dimension, induced by noise injected through the drive into a collective liquid state of light in a fast-gain laser. Using time-resolved spectroscopy, our measurements of transport along the synthetic frequency lattice reveal two distinct transitions, leading to three regimes of lattice occupation: a broad distribution at low noise, a Gaussian envelope at intermediate noise, and exponential localization at high noise levels.

While reminiscent of to studies of Anderson localization in photonic lattices, our work fundamentally differs by the absence of disorder along the lattice itself. Instead, we introduce a time-dependent environment, with an approach that is conceptually similar to spatial-to-temporal analog experiments~\cite{levi_hyper-transport_2012}. Yet, rather than observing transport enhancement, we demonstrate its suppression. This is a consequence of the nonlocal nonlinearity inherent to our non-equilibrium system, which influences the effect of a time-dependent environment and distinguishes our findings from studies of disorder-induced mobility transitions in linear regimes or with local nonlinearities. Furthermore, unlike implementations of temporal localization via temporal disorder, the nonlocal nonlinearity in our system, which maintains a constant total intensity, prevents localization in the time domain, confining all observed phenomena to the frequency domain where we detect them. Interestingly, our observations opens a relevant question concerning the timescales of temporal disorder which are relevant to the emergence of localization transitions. 

Finally, our work thus opens perspectives for understanding the role of temporal fluctuations on collective states of multimode light and offers potential pathways for their dynamical control through engineered temporal modulations, which could be relevant for applications requiring dynamical control of a laser's spectrum.

\section*{Acknowledgments}
This work was supported by the following: MIRAQLS: Staatssekretariat für Bildung, Forschung und Innovation SBFI (22.00182) in collaboration with EU (grant Agreement 101070700); Swiss National Science Foundation (212735); Innosuisse: Innovation Project 52899.1 IP-ENG (Agreement Number 2155008433 “High yield QCL Combs”); ETH Fellowship program: (22-1 FEL-46) (to A.D.)

\bigskip 
The data that support the findings of this article are openly available at~\cite{piciocchi_heckelmann_RC}.

\bigskip

D.P, I.H. and M.Ber. performed the measurements. D.P., M.S. ang C.J. performed the numerical simulations. I.H. conceptualized the experiment and processed the devices. M.Bec. grew the QCL wafer and consulted in the fabrication. D.P. wrote the manuscript draft and the Supplemental Material. D.P, I.H., A.D., O.Z. and J.F. collaborated on the interpretation of the results. A.D., C.J., and J.F. acquired and administrated the funding. All authors contributed to the review and editing of the final draft.

\newpage

\printbibliography[title={References}]

\newpage

\part*{Supplemental Material}
\setcounter{section}{0}

\section{Derivation of the Hamiltonian}\label{SI_ham}

Here, we detail the derivation of the Hamiltonian in Eq.\ref{eq:random} from the equation governing the electromagnetic field. Furthermore, we show how applying phase modulation with noise to a resonant drive results in a time-dependent gauge electric field on the synthetic lattice.

We begin with the current density driving the system, which, in the most general case relevant to this work, takes the form:
\begin{equation*}
    J(t) = J_0 + J_{\rm mod} \cdot\cos[(\Omega_{\rm res}+ \delta\Omega)t + \phi_{\rm e} N(t)].
\end{equation*}
Here, $J_0$ is the DC bias current, $J_m$ is the amplitude of the current modulation, $\Omega_{\rm res}$ is the cavity free-spectral range and $\delta\Omega$ is the detuning from the cavity resonance. $N(t)$ is a normalized random signal, and $\phi_{\rm e}$ is the applied phase excursion, which quantifies the noise level injected in the system. This current can also be expressed as: 
    \begin{equation*}
    J(t) = J_0 + J_{\rm mod} \cdot\cos\left[\int \Omega_{\rm inst}(t)\text{d}t\right], 
\end{equation*}
where the instantaneous frequency $\Omega_{\rm inst}$ is 
\begin{equation*}
    \Omega_{\rm inst} = \dv{[(\Omega_{\rm res}+ \delta\Omega)t + \phi_{\rm e} N(t)]}{t} = \Omega_{\rm res}+ \delta\Omega + \phi_{\rm e} \dv{N(t)}{t} =  \Omega_{\rm res}+ \delta\Omega + \Omega_{\rm N}(t).
\end{equation*}
This shows that the detuning from the cavity resonance becomes time-dependent upon phase modulation of the drive.

Next, we demonstrate how this time dependence translates to an effective electric field on a 1D lattice. This demonstration generalizes those reported in ~\cite{heckelmann_quantum_2023} and ~\cite{dikopoltsev_collective_2025}. We start from the cGLE equation describing field propagation in a reference frame co-propagating with the initial single mode:
\begin{equation*}
\begin{split}
     i\dot{E} &= i\left[g_0\left(1-\frac{I(t)}{I_{\rm sat}}\right) -\alpha\right] E +\left(\frac{i}{2}g_c - \frac{\beta}{2}\right)\nabla^2 E + \theta(t) \cdot 2C\cos\left[Kz-\left(\int \Omega_{\rm inst}(t)\text{d}t- \Omega_{\rm res}t\right)\right]E\\
     &= i\left[g_0\left(1-\frac{I(t)}{I_{\rm sat}}\right) -\alpha\right] E  +\left(\frac{i}{2}g_c - \frac{\beta}{2}\right)\nabla^2 E + \theta(t) \cdot 2C\cos\left[Kz-\delta\Omega \cdot t - \phi_{\rm e}N(t)\right]E.
\end{split}
\end{equation*}
Here, $g_0$ is the unsaturated gain, $I$ and $I_{\rm sat}$ are the intensity and saturation intensity, $\alpha$ is the total loss, $g_c$ is the gain curvature, $\beta$ is the dispersion and $K$ is the wavevector corresponding to the resonance frequency. The amplitude modulation is converted to phase modulation by the linewidth enhancement factor (LEF), and $2C$ is the resulting phase modulation coefficient. $\theta(t)$ is the Heaviside function.

Assuming the modulation is turned on at $t=0$ and focusing on $t\geq 0$, the linear and energy-conserving part of the equation becomes:
\begin{equation*}
    \dot{E} =  \left(i\frac{\beta}{2} + \frac{g_c}{2}\right)\nabla^2 E - i \cdot 2C\cos\left[Kz-v(t)\right]\cdot E,
\end{equation*}
where $v(t) =\delta\Omega t+\phi_{\rm e}N(t)$. This notation simplifies the expression and allows us to work with a generic form of time-dependent cavity detuning, potentially achieved through methods other than phase modulation. 

Expanding the field as $E(t)=\sum_n A_n(t)e^{-inKz}$ and substituting it into the previous equation yields:
\begin{equation*}
    \begin{split}
        \sum_n \dot{A}_n e^{-inKz} =& \left(i\frac{\beta}{2} + \frac{g_c}{2}\right)\sum_n(-inK)^2 A_n e^{-inKz} -2iC\cos\left[Kz-v(t)\right]\sum_n A_n e^{-inKz}\\ 
        =& -i\frac{(\beta-ig_c)K^2}{2}\sum_n n^2 A_n e^{-inKz} \\
        &-iC\sum_n A_n \left[ e^{-iv(t)}e^{-i(n-1)Kz} + e^{iv(t)}e^{-i(n+1)Kz} \right].
    \end{split}
\end{equation*}

Multiplying both sides by $e^{imKz}$, integrating over $z$ and defining $D=(\beta-ig_c)K^2 /2$, we obtain:
\begin{equation*}
    \dot{A}_m = -iDm^2 A_m -iC \left(e^{-iv(t)}A_{m+1} + e^{iv(t)}A_{m-1} \right).
\end{equation*}
By introducing the transformation $B_m = A_m e^{-imv(t)}$, we arrive at:
\begin{equation*}
    i\dot{B}_m = \left[D\cdot m^2+\Delta(T)\cdot m\right]B_m + C\left(B_{m+1} + B_{m-1}\right).
\end{equation*}
Here, we have switched to dimensionless quantities, with time $T=t/T_{\rm rt}$ measured in round-trip times and detuning in units of $\Omega_{\rm res}$: 
\begin{equation*}
    \Delta(T) = \frac{1}{\Omega_{\rm res}} \frac{\partial}{\partial T} v(T) = \frac{\delta\Omega}{\Omega_{\rm res}} + \frac{\phi_{\rm e}}{\Omega_{\rm res}}\dv{N(T)}{T} = \Delta + \frac{\Omega_{\rm N}(T)}{\Omega_{\rm res}}.
\end{equation*}

Introducing the creation and annihilation operators $a_m$ and $a^\dagger_m$ for photons in mode $m$, the Hamiltonian can be written as: 
\begin{equation*}
     H = \sum_m \left[D\cdot m^2+\Delta(T)\cdot m\right]a^\dagger_m a_m + C\left(a^\dagger_{m-1}a_m + a^\dagger_{m+1}a_m\right).
\end{equation*}

In the absence of detuning ($\Delta(T)=0$), this Hamiltonian reduces to the case presented by Heckelmann et al.~\cite{heckelmann_quantum_2023}. With a constant detuning ($N(t)=0\implies \Delta(T)=\Delta$) it simplifies to the scenario discussed by Dikopoltsev et al.~\cite{dikopoltsev_collective_2025}.  

\newpage 

\section{Envelope fit to identify transitions}\label{SI_fit}
\begin{figure}[h]
    \centering
    \includegraphics[width=0.7\linewidth]{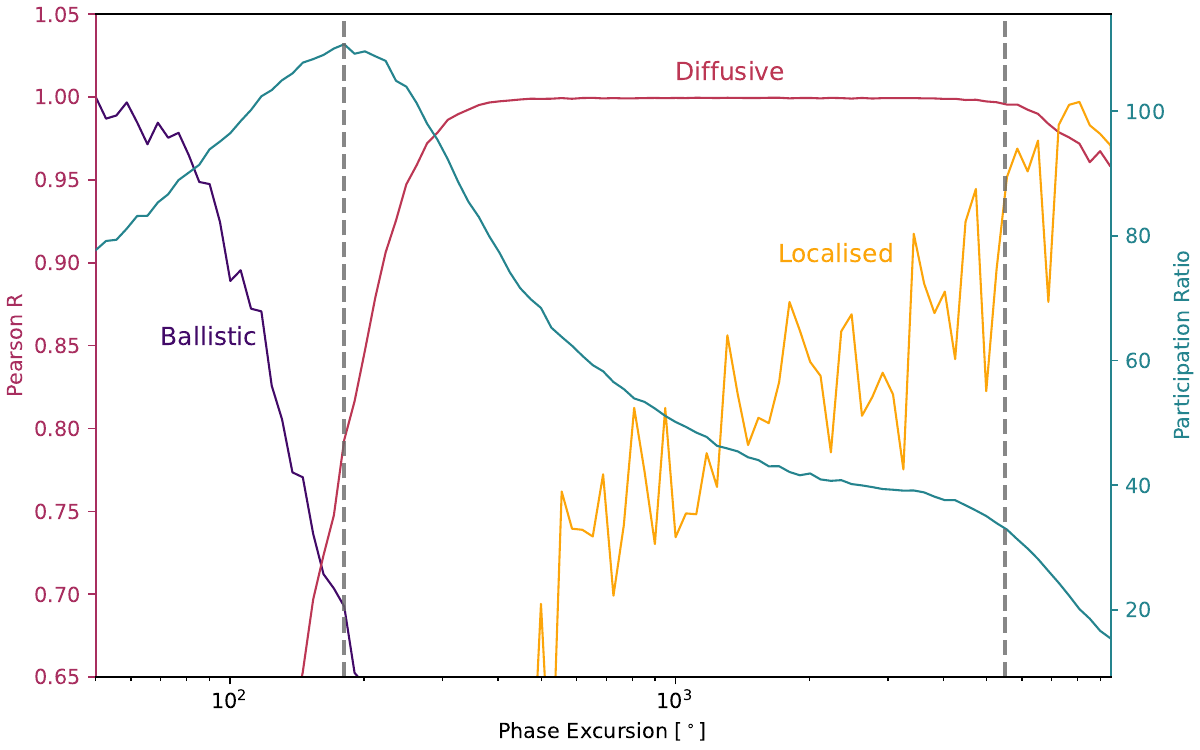}
    \caption{\textbf{Envelope shape transitions.} The participation ratio (blue curve) is shown alongside the Pearson R value for fits of three different spectral envelopes (red curves) as a function of the applied phase excursion.}\label{fig:envfit}
\end{figure}
In Fig.\ref{fig:transition} of the main text, the participation ratio is presented with examples of spectral envelopes to illustrate the correlation between changes in the participation ratio's trend and changes in the spectral shape. Here, we quantitatively demonstrate that the spectral envelope transitions occur at the levels of strength of the linear potential where the participation ratio's trend changes. For each $\phi_{\rm e}$value, we fit the measured average optical spectrum with a Quantum Walk-like frequency comb, a Gaussian, and an exponential function, assessing the fit quality using the Pearson R parameter. The results are shown together with the participation ratio in Fig.\ref{fig:envfit}. As $\phi_{\rm e}$ increases, the Quantum Walk fit, initially the best, progressively worsens as the participation ratio grows and the spectrum flattens, consistent with the main text. Around the maximum participation ratio, where its trend inverts, the Gaussian envelope becomes the best fit. This remains the case until the second abrupt change in the participation ratio's trend. At this point, the exponential fit becomes superior, although this fit is generally less robust, making the identification less immediate. 
\newpage

\section{Time-resolved setup}\label{SI_setup}
\begin{figure}[h]
    \centering
    \includegraphics[width=\linewidth]{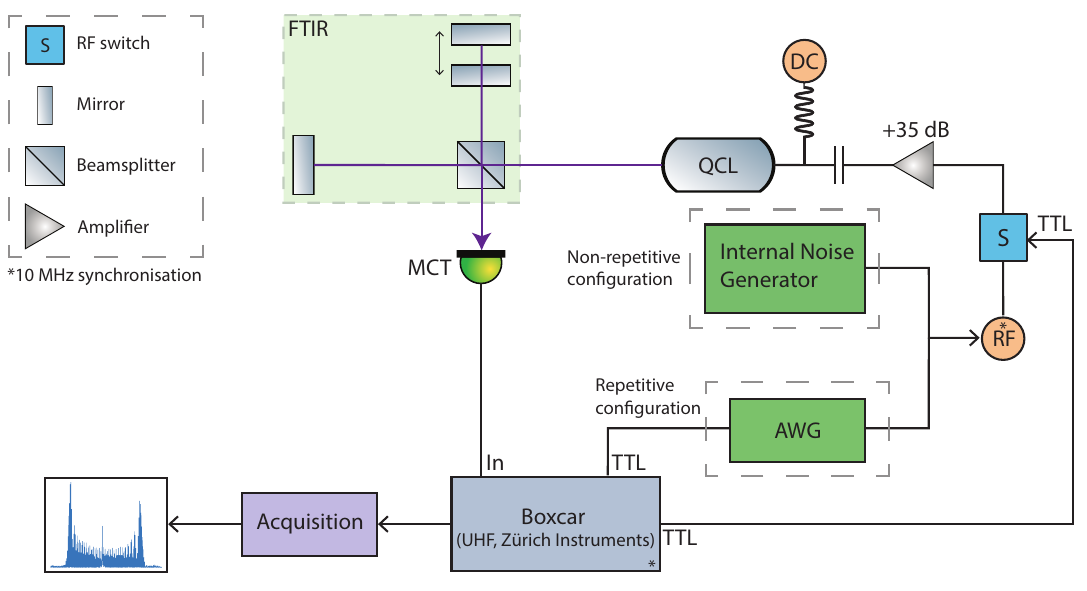}
    \caption{\textbf{Time-resolved measurement setup.} Schematic of the apparatus used to measure the time-resolved spectrum of the laser, both with a single noise trace (using the AWG) and averaging over many realizations (with the internal noise generator of the radio-frequency source).}\label{fig:timressetup}
\end{figure}
Fig.\ref{fig:timressetup} illustrates the experimental setup employed for the time-resolved measurement of the laser spectrum, providing information on the temporal evolution of the lattice occupation. The design is analogous to that reported in~\cite{heckelmann_quantum_2023}.

The time-dependent spectral reconstruction relies on boxcar averaging the signal from a Fourier Transform InfraRed (FTIR) spectrometer (Vertex 80 by Bruker), into which the QCL's light is directed. By progressively increasing the delay of the boxcar's window, we can reconstruct the spectrum at different times after the RF injection onset.

The QCL is continuously operated at a DC bias current of 1A (generated by a QCL2000 from Wavelength Electronics). To reconstruct the spectrum at specific times following the RF injection (provided by an SMF 100A from Rhode \& Schwarz), the latter is initiated using an RF switch (RFSPSTA0218G by RF-Lambda) triggered by the boxcar (UHF by Zurich Instruments) and amplified to 35dBm. This ensures that the modulation start time and the reference time for the boxcar delay are synchronized. The limitations imposed by the electronics, as discussed in~\cite{heckelmann_quantum_2023}, remain relevant. The triggering signal's rise time is 1ns, and the switch's rise time was characterized as 137ns, both significantly shorter than the timescales of the phenomena under investigation.

The photovoltaic Mercury-Cadmium-Telluride (MCT) detector (by Kolmar Technologies) at the FTIR output has a 20MHz 3dB bandwidth, allowing a 15dB SNR with a 10ns boxcar integration window.

The time step for increasing the boxcar delay was set to 10ns, and the trigger frequency to 50kHz. The latter results in a maximum measurable trace length of 10$\mu$s, as shown in Fig.\ref{fig:gauss} and \ref{fig:exploc} of the main text. To avoid aliasing, the FTIR's scanner velocity must be strictly below half the boxcar trigger frequency (in this case, 20kHz).

The RF signal was always centered at the cavity resonance frequency (15.7389GHz) and phase modulated with a Gaussian noise signal to produce the time-dependent instantaneous frequency. Three different phase modulation schemes were implemented.

For the integrated measurements, like the ones reported in Fig.\ref{fig:transition}, the AWG, Boxcar and Switch (S) were not used. The noise signal was produced by the internal noise generator of the RF source, and it was then used for phase modulation by the dedicated instrument function. In this simplified configuration, the MCT detector was directly connected to the acquisition. 

For the measurements in Fig.\ref{fig:gauss} and Fig.\ref{fig:exploc} (repetitive configuration), the noise signal was generated by an AWG (Agilent 33220A), triggered by the 50kHz boxcar trigger. The phase modulation was again performed by the dedicated function of the RF source. Within each 10$\mu$s active window initiated by the trigger, the AWG consistently produced the same noise trace, enabling the reconstruction of the spectrum's evolution for that specific trace.

In the third configuration, the RF source's internal noise generator was used, as for the measurements in Fig.\ref{fig:transition}. Consequently, during each active trigger window of the boxcar, the signal modulating the RF injection was always a different noise trace. This resulted in the spectrum for each boxcar delay value being averaged over $N=104$ different traces of the time-dependent injection frequency. The results of these measurements are shown in Fig.\ref{fig:rates}.

\newpage 

\section{Integration time}\label{SI_integration}
\begin{figure}[h]
    \centering
    \includegraphics[width=0.6\linewidth]{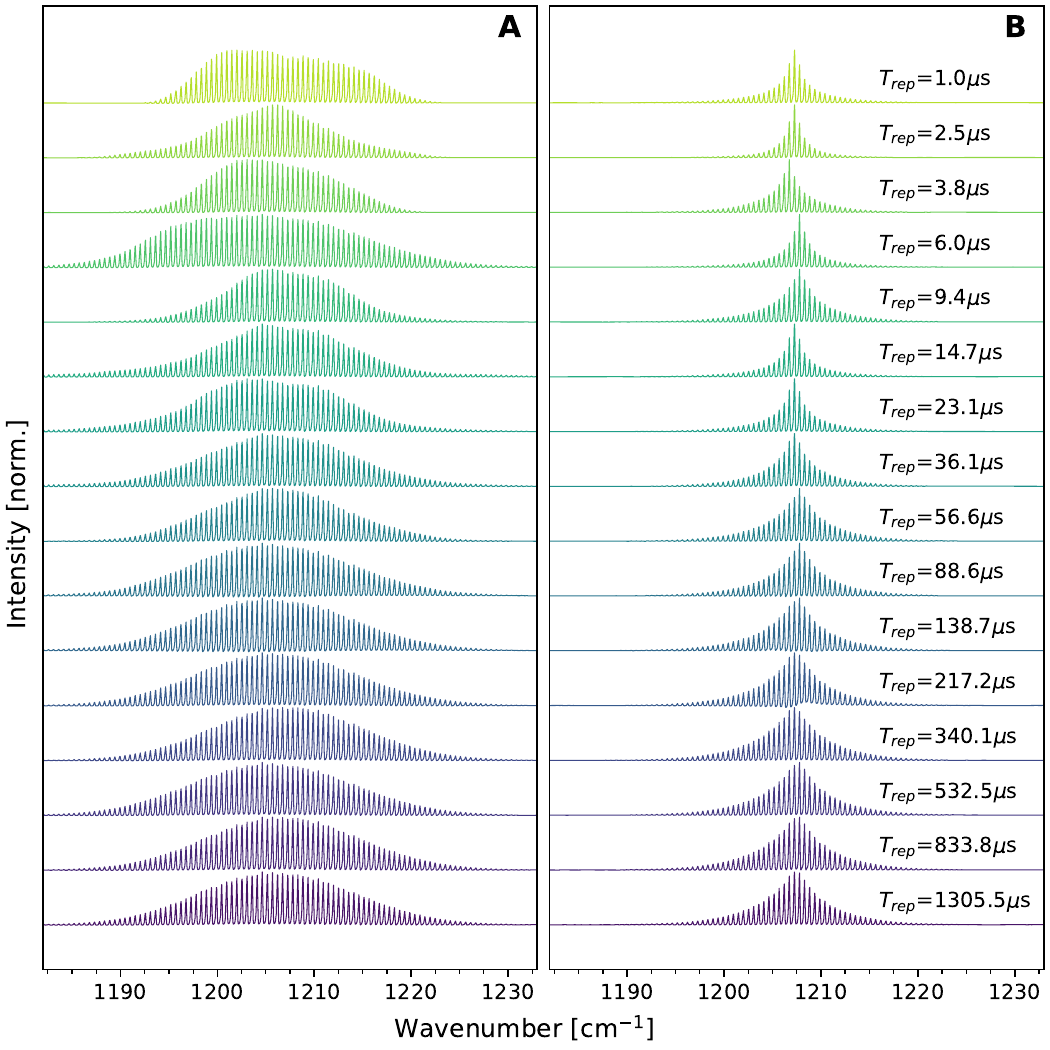}
    \caption{\textbf{Average spectra and integration time.} (A) Average spectra of the laser under noise injection with $\phi_e=1000^\circ/V$ measured with increasing integration time. (B) Average spectra of the laser under noise injection with $\phi_e=10000^\circ/V$ measured with increasing integration time.}
    \label{fig:integration}
\end{figure}
Fig.\ref{fig:integration} shows the measured average spectra for the phase excursion ($\phi_e$) values used in Fig.\ref{fig:gauss} (left, Fig.\ref{fig:integration}A) and Fig.\ref{fig:exploc} (right, Fig.\ref{fig:integration}B). Fig.\ref{fig:integration}A indicates that observing a Gaussian average spectrum requires an integration time of at least $\approx50\mu$s. Shorter integration times do not allow the system to sample a sufficient number of different configurations, resulting in average spectra with different shapes. In contrast, Fig.\ref{fig:integration}B shows that even at short integration times, the average spectrum exhibits an exponentially localized shape. This is consistent with Fig.\ref{fig:exploc}, which already shows exponential localization in all temporal slices. The broadening of the exponential at longer integration time is solely due to the limited number of points generated by the AWG for a noise signal with a duration equal to the integration time. As the time trace lengthens, the finite number of points effectively downsamples the noise, altering its frequency content and thus affecting the spectrum.  
\newpage

\section{Numerical Simulations}\label{SI_sims}
\begin{figure}[h!]
  \begin{minipage}{0.49\textwidth}
    \centering
    \includegraphics[width=\linewidth]{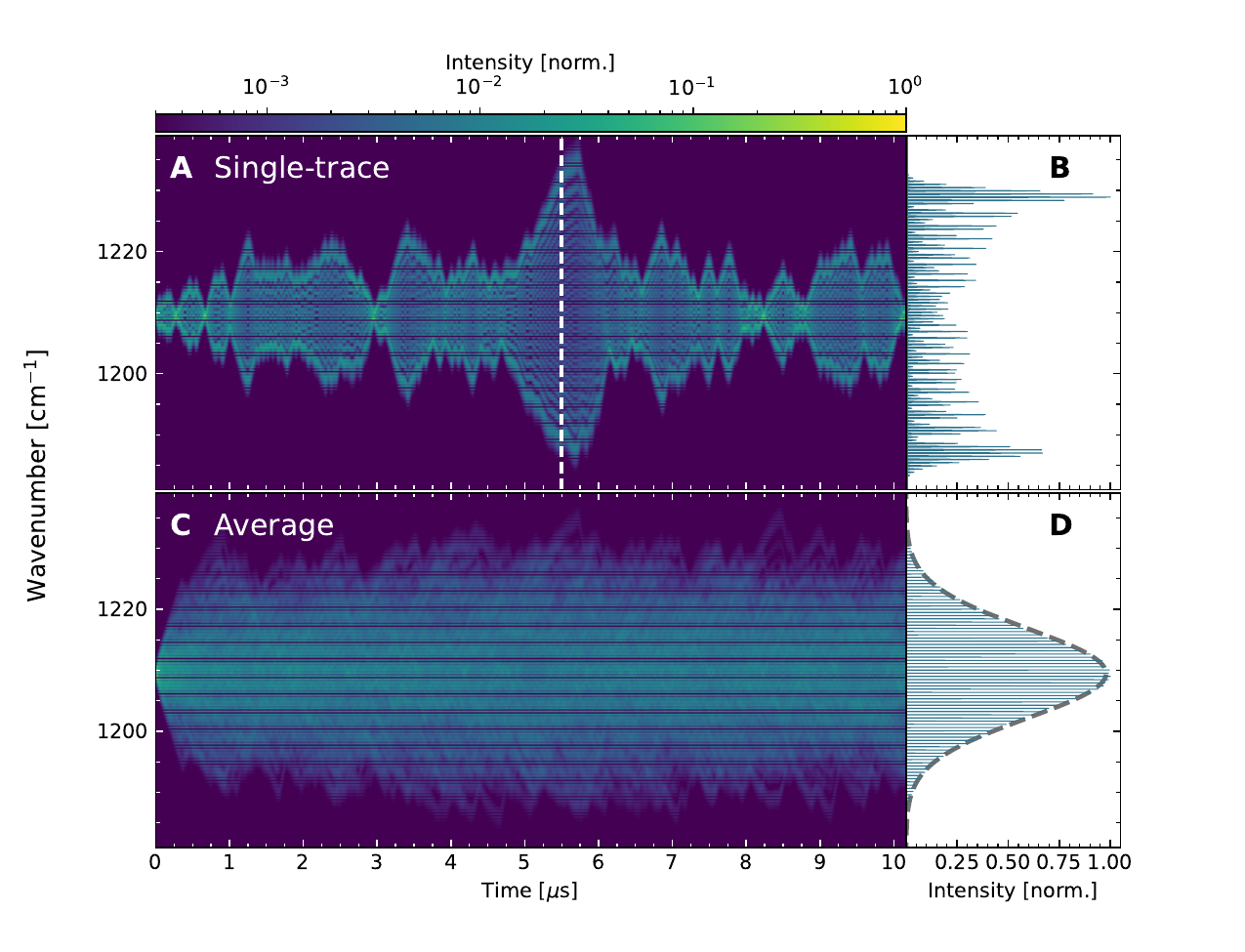}
    \caption{\textbf{Time-evolution simulation in the Gaussian regime.} (A) Simulation of the time evolution of the laser spectrum under noise injection for a single noise trace. (B) Spectrum of the laser at a chosen time during the evolution, indicated by the white dashed line. (C) Time evolution of the laser spectrum, averaged over many realizations of a noise time trace of the same duration. (D) Average of all the simulated spectra between $t=1\mu$s and $t=10\mu$s, plotted with a Gaussian fit on the experimental data in Fig.\ref{fig:gauss} (dashed gray curve).}
    \label{fig:simgauss}
  \end{minipage}\hfill%
  \begin{minipage}{0.49\textwidth}
    \centering
    \includegraphics[width=\linewidth]{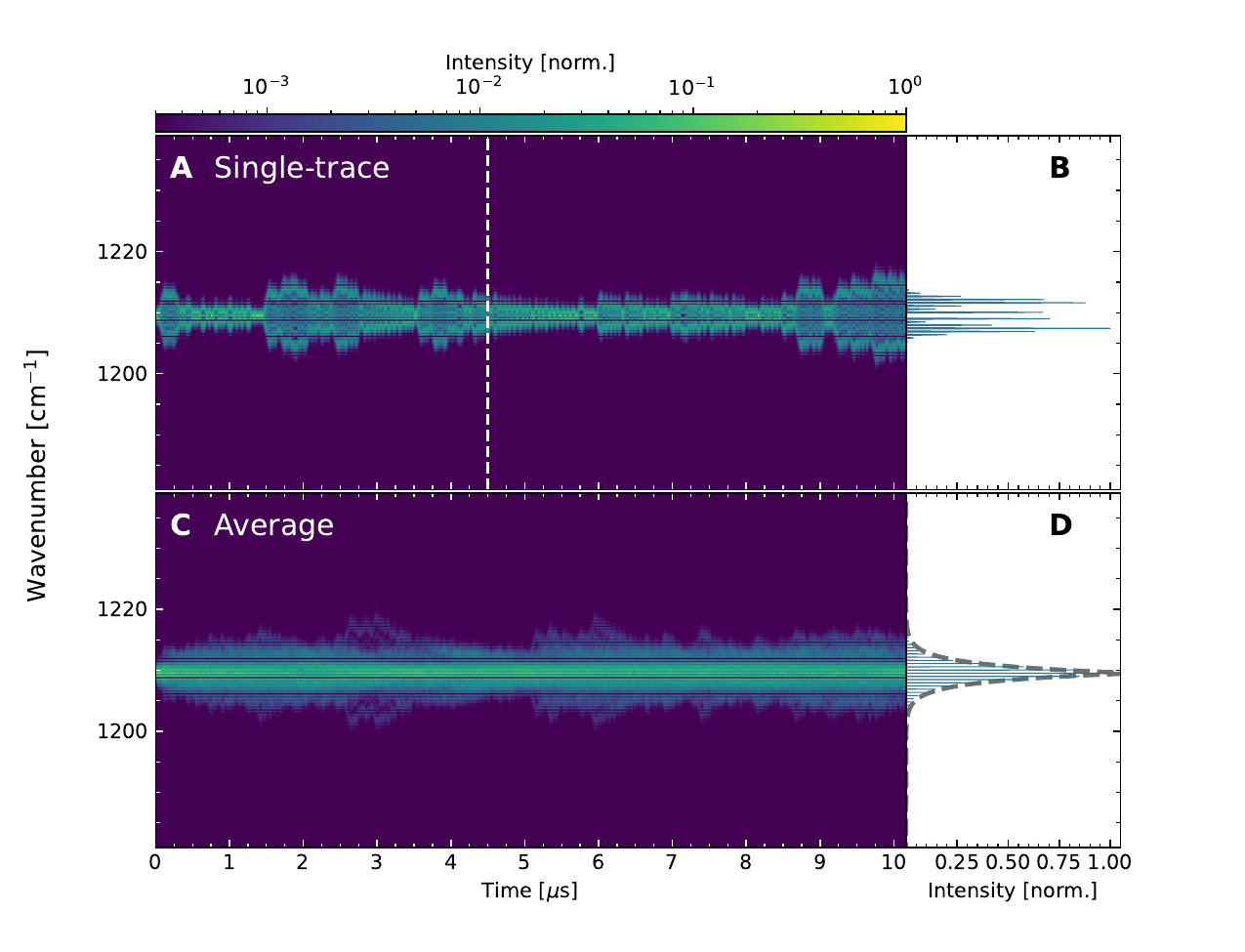}
    \caption{\textbf{Time-evolution simulation in the localized regime.} (A) Simulation of the time evolution of the laser spectrum under noise injection for a single noise trace. (B) Spectrum of the laser at a chosen time during the evolution, indicated by the white dashed line. (C) Time evolution of the laser spectrum, averaged over many realizations of a noise time trace of the same duration. (D) Average of all the simulated spectra between $t=1\mu$s and $t=10\mu$s, plotted with an exponentially localized fit on the experimental data in Fig.\ref{fig:exploc} (dashed gray curve).}
    \label{fig:simexp}
  \end{minipage}
\end{figure}

Fig.\ref{fig:simgauss} and Fig.\ref{fig:simexp} present numerical simulation results with parameters matching the noise injected during the measurements shown in Fig.\ref{fig:gauss} and Fig.\ref{fig:exploc}, respectively.

These results were obtained by solving the complex Ginzburg-Landau equation (Eq.\ref{eq:cGLE}) using a split-step method. The calculation of the field's derivative at each iteration (corresponding to a time step) is divided into two parts (see Supplementary Information in~\cite{heckelmann_quantum_2023}). First, the field is Fourier transformed, and the part of the evolution operator diagonal in the frequency domain, representing dispersion and gain curvature, is applied. Then, the field is transformed back to real space, and the remaining part of the evolution operator, accounting for gain saturation and modulation, is applied. The modulation includes a term induced by the RF signal's phase modulation, which creates the time-dependent electric gauge field along the synthetic lattice: $\theta(t) \cdot 2C\cos\left[Kz-\delta\Omega \cdot t - \phi_{\rm e}N(t)\right]E$.

In all simulations, the $N(t)$ trace used was the same as in the experiment, sampled using an oscilloscope (Teledyne LeCroy HDO6104, 12 bit, 500MHz bandwidth). This trace was then digitally low-pass filtered during the simulation to remove high-frequency noise and match the maximum bandwidth of the AWG used experimentally. Finally, the trace was rescaled to a $\phi_e$ value matching the experimental one. 

In the case of Fig.\ref{fig:simgauss}, where the average occupation (or laser spectrum) is Gaussian, a simple rescaling of $\phi_e$ sufficed to reproduce the experimental results with high quantitative accuracy. We observe that the simulated spectrum also exhibits continuous time evolution (Fig.\ref{fig:simgauss}A), and no individual spectrum resembles a Gaussian (Fig.\ref{fig:simgauss} A). The Gaussian spectrum only emerges upon averaging over a long time trace (or equivalently, over many realizations of a fixed-duration noise trace), as shown in Fig.\ref{fig:simgauss}C and D. 

The situation differed for the exponentially localized spectra shown in Fig.\ref{fig:simexp}. Here, achieving quantitative agreement required adding a fixed detuning of the injection frequency in addition to rescaling the trace. However, as Fig.\ref{fig:simexp} A and B illustrate, the simulations could not reproduce individual spectra with exponentially decaying shapes during the time evolution. Quantitative agreement with the measured data was only achieved upon averaging, as shown in Fig.\ref{fig:simexp}.

To further investigate the exponentially decaying spectra, we conducted extensive numerical simulations based on the Maxwell-Bloch equations in the rotating-wave and slowly-varying envelope approximations~~\cite{allen_optical_1987,jirauschek_optoelectronic_2019}. The Maxwell-Bloch equations constitute a non-adiabatic generalization of the aforementioned cGLE~~\cite{opacak_frequency_2021}. In this model, the evolution of the complex field envelope $E$ is governed by the propagation equation~~\cite{jirauschek_optoelectronic_2019}
\begin{equation*}
     v_\mathrm{g} \frac{\partial E}{\partial t} + \frac{\partial E}{\partial z}  = f - \alpha E -  \frac{i \beta}{2} \frac{\partial^2 E}{\partial t^2} \, . 
\end{equation*}
Here, $v_\mathrm{g}$ denotes the group velocity and $f$ represents the polarization caused by the active region, which can be directly computed from the Bloch equations. For numerical simulation, the propagation equation is discretized using the Risken-Nummedal finite-difference scheme~~\cite{risken_self-pulsing_1968}, where the dispersion is treated separately via a Strang splitting method. To model the ring geometry of the device, periodic boundary conditions are imposed onto the field envelope, i.e., $E(t,0)=E(t,L)$.
Moreover, the dynamics of the active region is captured by an effective two-level description. The state of the two-level quantum system is described by the population inversion $w$ and the coherence $\eta_{21}$, which obey the Bloch equations~~\cite{jirauschek_optoelectronic_2019}
\begin{align*}
    \partial_t \eta_{21} &= i \Delta \eta_{21} - i \frac{d_{21}}{2\hbar} \, w \, E - \gamma_2 \, \eta_{21} \, , \\
    \partial_t w &=  i \frac{d_{21}}{\hbar}  \left( \eta_{21}^{*} E - \eta_{21} E^{*} \right) - \gamma_1 ( w - w_\mathrm{eq} ) \, .
\end{align*}
Here, $d_{21}$ denotes the dipole moment, $\gamma_2$ and $\gamma_1$ are the dephasing and the gain recovery rates, and $\Delta = \omega_\mathrm{c} - \omega_{21}$ represents the detuning between the center frequency $\omega_{c}$ and the resonance frequency $\omega_{21}$ of the two-level system. Input parameters for the simulations were self-consistently extracted from ensemble Monte Carlo carrier transport simulations of the quantum active region, that were subsequently mapped to a reduced two-level description. The Bloch equations are numerically treated using the $3^\mathrm{rd}$ order Adams–Bashforth method~~\cite{jirauschek_optoelectronic_2019}. We incorporate the RF modulation by coupling solutions of the transmission line equations for a ring waveguide to the Maxwell-Bloch model~~\cite{jirauschek_theory_2023}.

\begin{figure}[h]
    \centering
    \includegraphics[width=0.6\linewidth]{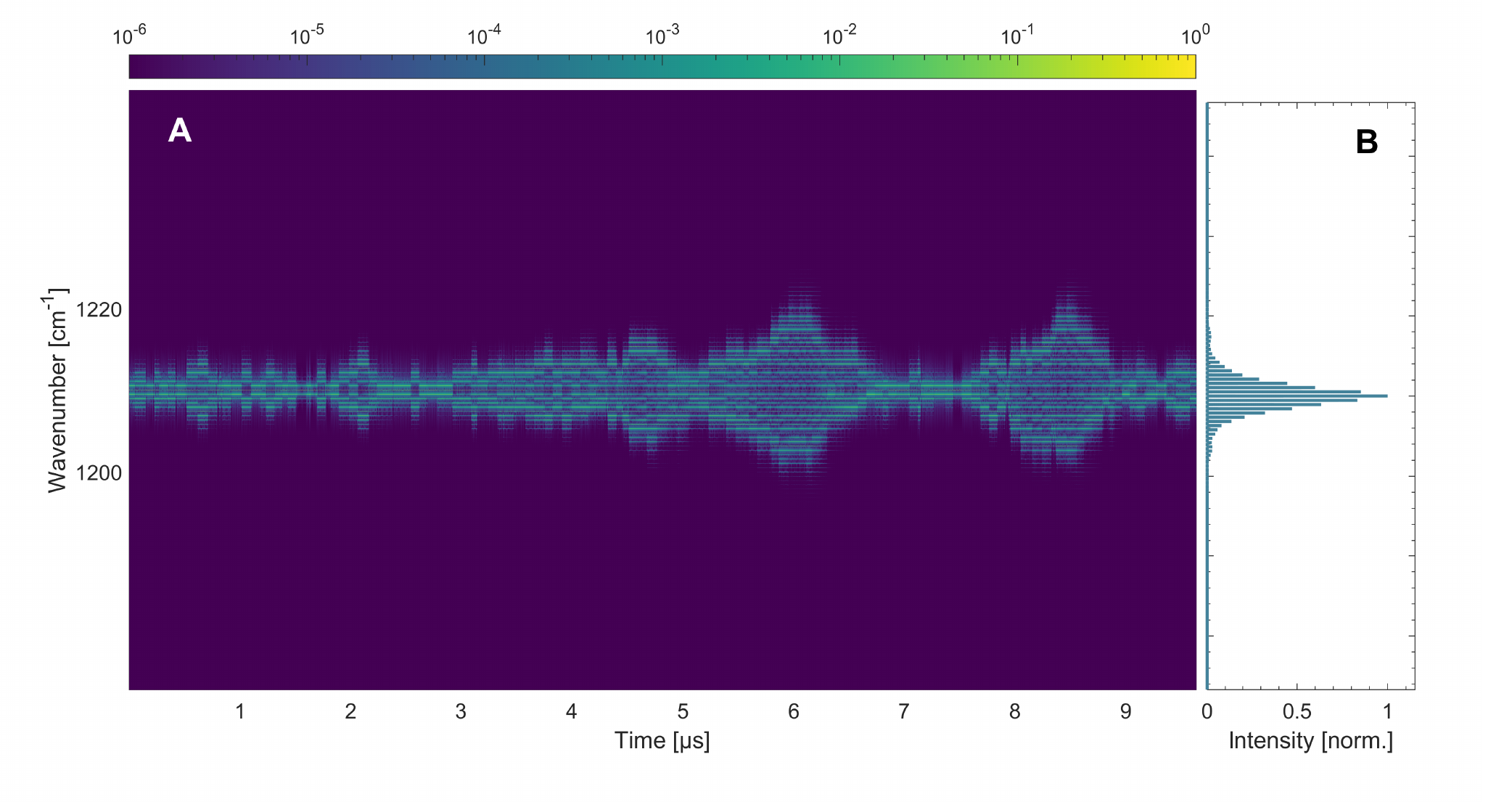}
    \caption{\textbf{Time-evolution simulation from the Maxwell-Bloch model  in the localized regime.} (A) Simulation of the time evolution of the laser's spectral intensity under noise injection for a single noise trace, and (B) corresponding intensity spectrum.}
    \label{fig:MB-simulation}
\end{figure}

Simulation results for the exponential regime, obtained from this model, are displayed in Fig.~\ref{fig:MB-simulation}. Here, the averaged intensity spectrum shows a exponential decaying envelope as in the experimental data (see Fig.~\ref{fig:transition}). In the simulation, the ring laser is resonantly injected with an RF wave. As in previous simulations, the noise trace $N(t)$ included into the injection signal corresponds to the one used in experiment, and $\phi_e$ is rescaled to match the experiment.

\subsection{Comparison with the slow gain}\label{SI_sims_sg}
\begin{figure}[t]
\centering
\includegraphics[width=0.65\linewidth]{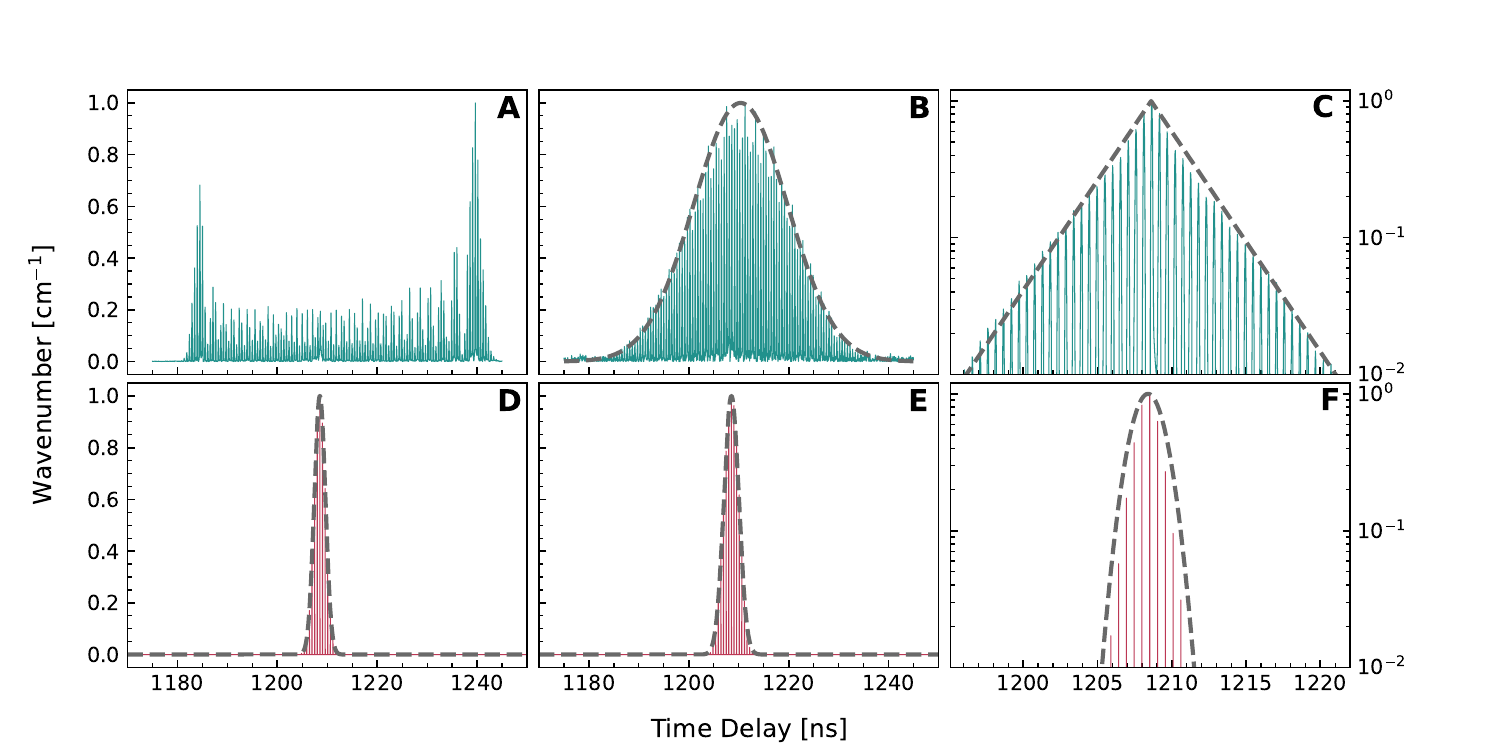}
\caption{\textbf{Average spectrum of fast and slow gain systems.} Average optical spectrum of the system with fast (top row) and slow (bottom row) gain. A, D show the case where no phase modulation is applied. B, E correspond to a case where phase modulation produces a Gaussian envelope in the fast gain system. C, F correspond to an exponential envelope condition for the fast gain system.}\label{fig:avgsg}
\end{figure}
In the main text, we argued that the fast gain of the QCL, acting as a nonlinear term in the equation describing the synthetic 1D lattice, plays a crucial role in enabling the observation of both Quantum Walk dynamics~\cite{heckelmann_quantum_2023} and the transitions described in our work. Here, we support this by presenting simulations of slow gain systems driven under the same conditions as the fast gain systems. This involves applying the same time-dependent force but replacing the nonlinear term as follows:
\begin{equation*}
    F_{\rm NL, fast}(t) = g_0\left(1-\frac{I(t)}{I_{\rm sat}}\right) \quad \to \quad 
    F_{\rm NL, slow}(t) =g_0\left(1-\frac{\langle I(t)\rangle}{I_{\rm sat}}\right)
\end{equation*}

A slow gain medium does not respond to the instantaneous intracavity intensity during a round-trip but saturates based on the average intracavity field intensity over that round-trip. Consequently, slow gain media lack the intensity fluctuation smoothing property that enforces quasi-CW operation in QCLs.
\begin{figure}[ht]
    \centering
    \includegraphics[width=0.7\linewidth]{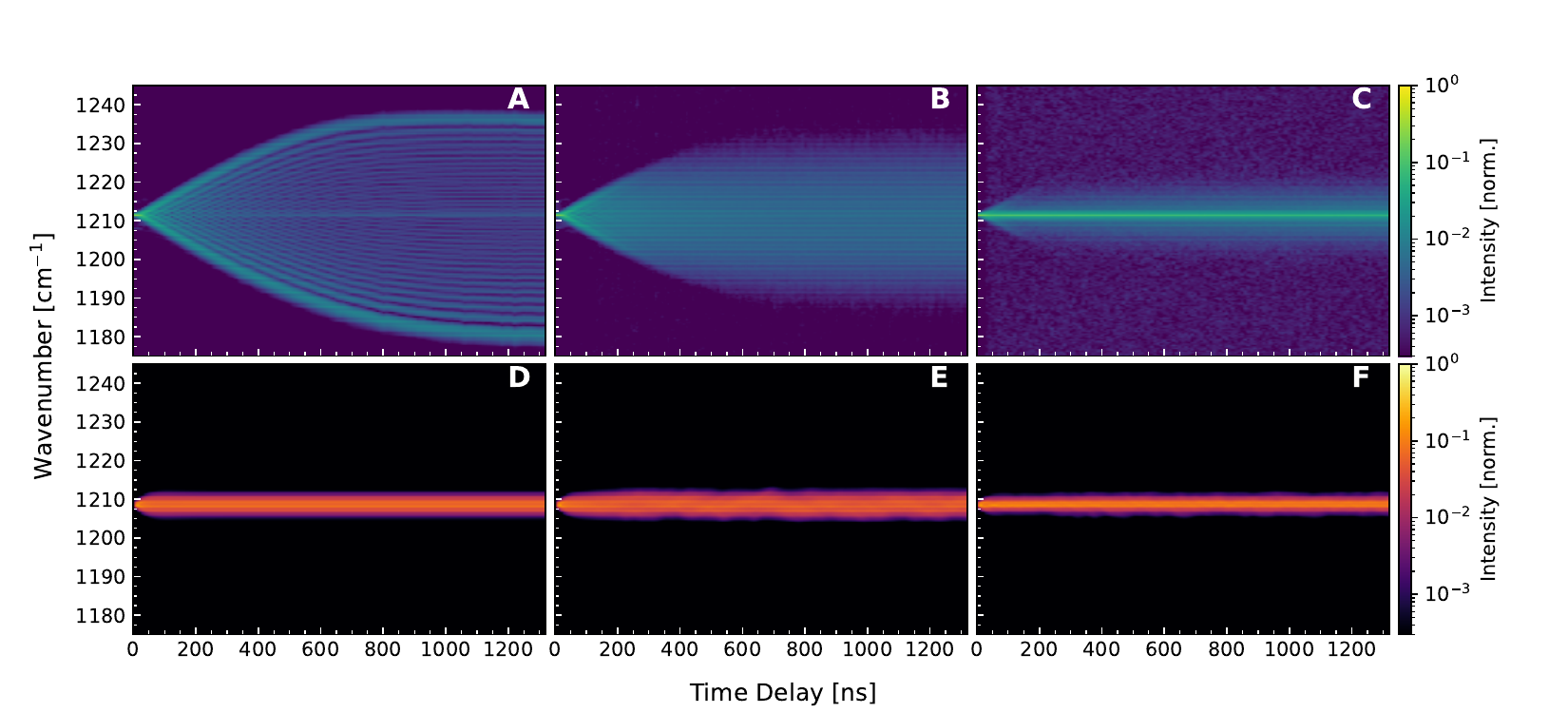}
    \caption{\textbf{Temporal dynamics of fast and slow gain systems.} Time-resolved optical spectrum of the system with fast (top row) and slow (bottom row) gain. A, D show the case where no force is applied. B,C,D and E show the case where a time-dependent drive is used, showcasing different realizations of the drive history. Cases corresponding to a Gaussian (B,E) or exponentially localized (C,F) spectrum are shown.}\label{fig:mapsg}
\end{figure}

This distinction is relevant for understanding how the steady state of the quantum walk dynamics is a high-order supermode of the coupled system. The quasi-CW nature of this state implies it must be broad in time. According to the time-bandwidth product theorem, a system in the 0th order Gaussian supermode that is broad in time must be spectrally narrow, and vice versa. However, this does not hold for the QCL quantum walk comb, which is both spectrally broad and quasi-CW, residing in a high-order supermode~\cite{heckelmann_quantum_2023} that is broad in both time and frequency and exhibits a frequency-modulated comb behavior. The steady-state spectra of identical systems differing only in the nonlinearity are shown in Fig.\ref{fig:avgsg} A and D. The fast gain system displays the broad quantum walk spectrum, while the slow gain system settles into a narrow Gaussian state.

The fast gain's role in enforcing a broad spectrum (lattice occupation) is also essential for observing a response to changes in the driving signal that enable nearest-neighbor coupling between the modes (lattice sites). Panels B, C, E, F of Fig.\ref{fig:avgsg} demonstrate that while the fast-gain system exhibits transitions between different regimes (as discussed in the main text), the slow-gain system (reproduced here with numerical simulations) shows only minor bandwidth changes, maintaining Gaussian envelope spectra. A similar effect is seen in the comparison of the time-resolved dynamics of the systems, presented in Fig.\ref{fig:mapsg}.
\newpage

\section{Averaged time-resolved dynamics}\label{SI_avg_tres}
\begin{figure}[h]
\centering
\includegraphics[width=0.7\linewidth]{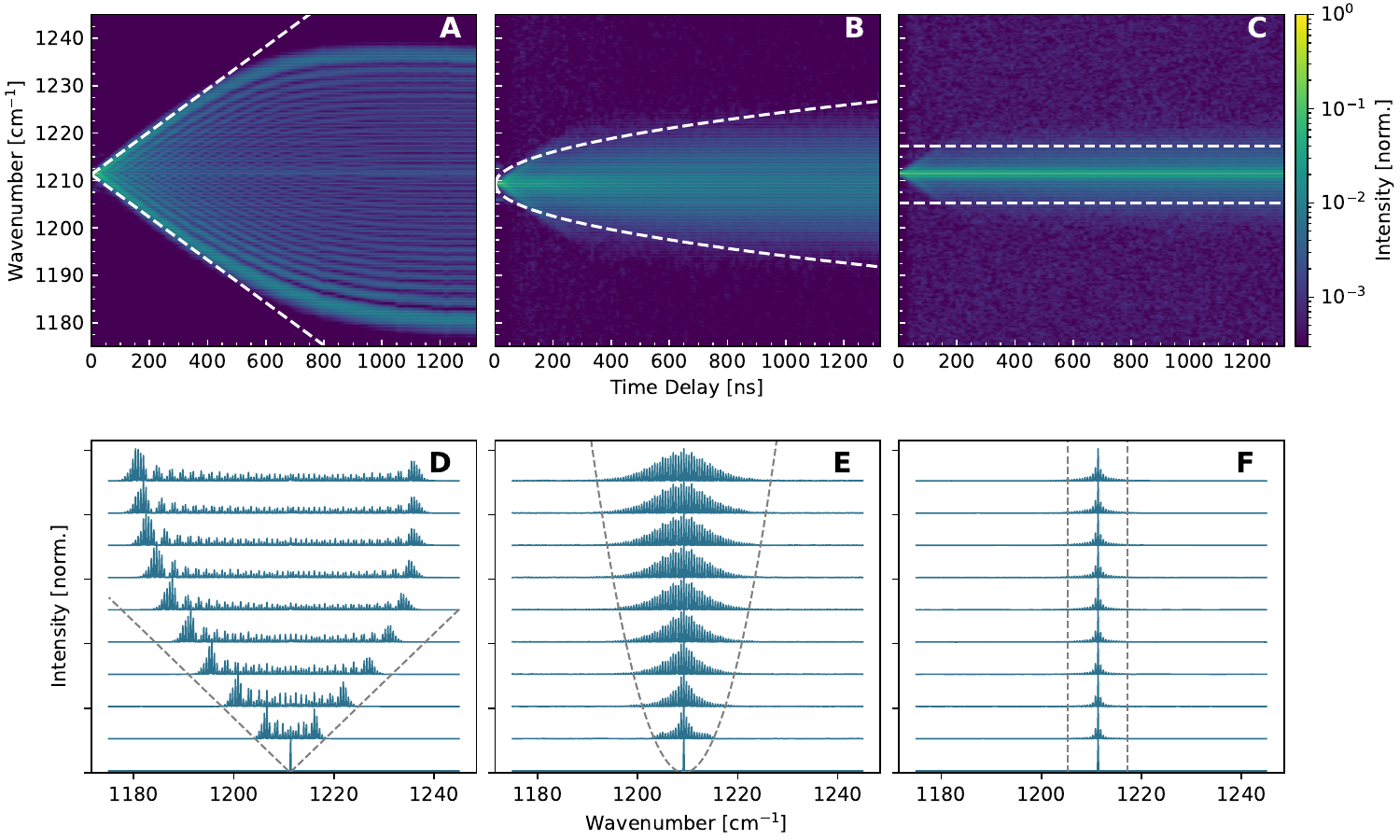}
\caption{\textbf{Averaged time-resolved measurements.} (A-C) Time-resolved measurement of the spectrum for different applied force levels. One case is shown for each relevant situation: ballistic (A), Gaussian (B) and exponentially localized (C). (D-F) Some spectra at increasing times. The dashed lines indicate the ballistic, diffusive, and arrested expansion trends.}\label{fig:avgtrse}
\end{figure}
Fig.\ref{fig:avgtrse} presents the same time-resolved expansion maps as Fig.\ref{fig:rates} in the main text, where the occupation at each time point is averaged over multiple realizations of the applied noise.

Panels A and D show the case without applied force, leading to quantum random walk dynamics. The linear expansion trend of the lattice occupation is indicated by the dashed line. Panels B and E show the intermediate regime with Gaussian envelopes, exhibiting a sub-ballistic expansion trend (that is, a power law $t^k$ with $k<1$). Finally, panels C and F illustrate the case of exponentially decaying envelopes, where the spectrum does not expand beyond the first few populated modes (highlighted by the vertical dashed lines).

\end{document}